\documentclass[12pt]{article}
\usepackage{amssymb}
\usepackage{amsmath}
\usepackage[unicode,bookmarks,bookmarksopen,bookmarksopenlevel=2,colorlinks,linkcolor=blue,citecolor=green]{hyperref}

\input{tcilatex}
\begin{document}

\title{\textbf{Dispersionful Version of WDVV Associativity System}}
\author{Maxim V. Pavlov$^{1,2}$, Nikola M. Stoilov$^{3,4}$ \\
$^{1}$Sector of Mathematical Physics,\\
Lebedev Physical Institute of Russian Academy of Sciences,\\
Leninskij Prospekt 53, 119991 Moscow, Russia\\
$^{2}$Department of Applied Mathematics,\\
National Research Nuclear University MEPHI,\\
Kashirskoe Shosse 31, 115409 Moscow, Russia\\
$^{3}$Max Planck Institute for Dynamics and Self-Organization,\\
37077 G\"ottingen, Germany\\
$^{4}$Mathematisches Institut, Georg-August-Universit\"at G\"ottingen,\\
Bunsenstra{\ss }e 3-5, 37073 G\"ottingen, Germany}
\date{\today}
\maketitle

\begin{abstract}
B.A. Dubrovin proved that remarkable WDVV associativity equations are
integrable systems. In a simplest nontrivial three-component case these
equations can be written as a nondiagonalizable hydrodynamic type system
equivalent to a symmetric reduction of the three wave interaction and to the
matrix Hopf equation. Then E.V. Ferapontov and O.I. Mokhov found a local
Hamiltonian structure. Finally E.V. Ferapontov, C.A.P. Galv\~{a}o, O.I.
Mokhov, Ya. Nutku found a second local Hamiltonian structure. Both local
Hamiltonian structure are homogeneous of first and third order
(respectively) of Dubrovin--Novikov type.

In our paper we suggest a special scaling procedure for independent
variables applicable for homogeneous nonlinear PDE's, which allows to
incorporate an auxiliary parameter $\epsilon $, such that a corresponding
\textquotedblleft intermediate\textquotedblright\ system possesses two
remarkable limits: a high-frequency limit ($\epsilon \rightarrow \infty $)
back to the original system and a dispersionless limit ($\epsilon
\rightarrow 0$) which yields diagonalizable integrable hydrodynamic type
system. This means that our procedure allows to transform a homogeneous
third order local Hamiltonian structure to non-homogeneous of third order.
Thus we create an integrable hierarchy equipped by a pair of local
Hamiltonian structures, which (both of them) possess a dispersionless limit.
Also we show that this bi-Hamiltonian diagonalizable hydrodynamic type
system possesses at least two different dispersive integrable extensions (in
a framework of B.A. Dubrovin's approach).
\end{abstract}

\textit{On the occasion of Boris Dubrovin's 65th birthday}

\tableofcontents

keywords: associativity equations, integrable dispersive systems,
dispersionless limit, high-frequency limit, bi-Hamiltonian structure,
hydrodynamic type system.\newpage 

\section{Introduction}

Plenty of integrable systems possess a dispersionless limit, which is a
semi-Hamiltonian hydrodynamic type system. Most well-known examples are: the
Korteweg de Vries equation, the Kaup--Boussinesq system, the Boussinesq
equation. However some other integrable systems have no such a
dispersionless limit (at least nobody knows how to derive it). Examples: the
Bonnet equation (also known as the Sin-Gordon equation), the
Krichever--Novikov equation, the Landau--Lifshitz equation.

In this paper we start from the hydrodynamic type system%
\begin{equation}
a_{y}^{1}=a_{z}^{2},\text{ \ \ }a_{y}^{2}=a_{z}^{3},\text{ \ \ }%
a_{y}^{3}=[(a^{2})^{2}-a^{1}a^{3}]_{z}.  \label{wdvv}
\end{equation}%
This system is nondiagonalizable, i.e. is not semi-Hamiltonian, but is
integrable by the inverse scattering transform. This system was derived as a
single equation ($a^{1}=f_{zzz},a^{2}=f_{yzz},a^{3}=f_{yyz}$)%
\begin{equation}
f_{yyy}=f_{yzz}^{2}-f_{zzz}f_{yyz}  \label{sing}
\end{equation}%
by B.A. Dubrovin (see detail in \cite{Dubr}) and rewritten in the above
hydrodynamic form by E.V. Ferapontov and O.I. Mokhov (see detail in \cite{FM}%
). A standard approach established by B.A. Dubrovin for reconstruction of
higher dispersive corrections to this case is inapplicable, because a
dispersionless limit must be a semi-Hamiltonian hydrodynamic type system.
Nevertheless we were able to construct a new integrable system, which we
call the \textquotedblleft intermediate\textquotedblright\ dispersive
system. This intermediate dispersive system essentially depends on an extra
parameter $\epsilon $ such that: if $\epsilon \rightarrow \infty $, in this
high-frequency limit one can obtain the above non-diagonalizable
hydrodynamic type system; if $\epsilon \rightarrow 0$, in this
dispersionless limit one can obtain a semi-Hamiltonian hydrodynamic type
system.

The main advantage of our construction is that we just re-scaled the
independent variable $z\rightarrow z(x,\epsilon )$ in a linear equation of
third order, which is a first \textquotedblleft half\textquotedblright\ of
the Lax pair associated with the above non-diagonalizable hydrodynamic type
system. This non-diagonalizable hydrodynamic type system possesses a
bi-Hamiltonian structure (see detail in \cite{FGMN}). In this paper we show
that the intermediate dispersive system also has a bi-Hamiltonian structure
as well as its dispersionless limit.

The structure of the paper is as follows. In Section~\ref{sec:biham} we
briefly describe a bi-Hamiltonian formulation of the WDVV associativity
system and discuss a Theorem about a relationship between flat coordinates
and a momentum for a local Hamiltonian structure of Dubrovin--Novikov type
of first order \cite{DN}. In Section~\ref{sec:trans} we construct a
one-parametric transformation from the first \textquotedblleft
half\textquotedblright\ of the Lax pair (determining the WDVV associativity
system) to another first \textquotedblleft half\textquotedblright\ of a Lax
pair, which determines a new integrable dispersive system, which depends on
an arbitrary parameter. In Section~\ref{sec:inter} we present a procedure
for construction of local Hamiltonian structures of Dubrovin--Novikov type.
In Subsection \ref{subsec:density} we expand a generating function of
conservation law densities and extract two quadratic relationships between
conservation law densities, which should be identified as flat coordinates
and momenta. In Subsection \ref{subsec:evolu} we find a non-evolutionary
compact form for the intermediate dispersive system, which allows to
consider a high-frequency limit. We construct a bi-Hamiltonian structure and
present two corresponding local Lagrangian representations. In Subsection %
\ref{subsec:wdvv} we demonstrate that a high-frequency limit of this
intermediate dispersive system is precisely the WDVV associativity system (%
\ref{wdvv}). In Section \ref{sec:expand} we slightly re-scale the
independent variable \textquotedblleft $x$\textquotedblright\ and the
spectral parameter $\lambda $. Then all conservation law densities become of
hydrodynamic type in a dispersionless limit. In Subsection \ref%
{subsec:disperless} we consider a semi-Hamiltonian system, which follows
from a dispersionless limit. We construct its bi-Hamiltonian structure and
show that this dispersionless system coincides with a dispersionless limit
of the remarkable Yajima--Oikawa system. Thus this bi-Hamiltonian
hydrodynamic type system possesses at least two different dispersive
integrable extensions. In Subsection \ref{subsec:lagran} we find a
non-evolutionary compact form for the intermediate dispersive system, which
allows to consider a dispersionless limit. In Section \ref{sec:oikawa} we
briefly consider the Yajima--Oikawa system, its bi-Hamiltonian structure and
a dispersionless limit. In Section \ref{sec:another} we discuss a similar
transformation for the second \textquotedblleft half\textquotedblright\ of
the Lax pair (determining the WDVV associativity system). Finally in
Conclusion

\section{Bi-Hamiltonian Structure of WDVV Associativity System}

\label{sec:biham}

The remarkable WDVV associativity equation (\ref{sing}) written in an
equivalent hydrodynamic form (\ref{wdvv}) admits a bi-Hamiltonian structure
(see detail in \cite{FGMN})%
\begin{equation*}
a_{y}^{i}=A_{1}^{ij}\frac{\delta \mathbf{H}_{2}}{\delta a^{j}}=A_{2}^{ij}%
\frac{\delta \mathbf{H}_{1}}{\delta a^{j}},
\end{equation*}%
where two compatible local Hamiltonian operators $\hat{A}_{1}$ and $\hat{A}%
_{2}$ are%
\begin{gather}
\hat{A}_{1}=%
\begin{pmatrix}
-\frac{3}{2}\partial _{z}^{{}} & \frac{1}{2}\partial _{z}^{{}}a^{1} & 
\partial _{z}^{{}}a^{2} \\ 
\frac{1}{2}a^{1}\partial _{z}^{{}} & \frac{1}{2}(\partial
_{z}^{{}}a^{2}+a^{2}\partial _{z}^{{}}) & \frac{3}{2}a^{3}\partial
_{z}^{{}}+a_{z}^{3} \\ 
a^{2}\partial _{z}^{{}} & \frac{3}{2}\partial _{z}^{{}}a^{3}-a_{z}^{3} & 
[(a^{2})^{2}-a^{1}a^{3}]\partial _{z}^{{}}+\partial
_{z}^{{}}[(a^{2})^{2}-a^{1}a^{3}]%
\end{pmatrix}%
,  \label{i} \\
\hat{A}_{2}=%
\begin{pmatrix}
0 & 0 & \partial _{z}^{3} \\ 
0 & \partial _{z}^{3} & -\partial _{z}^{2}a^{1}\partial _{z} \\ 
\partial _{z}^{3} & -\partial _{z}a^{1}\partial _{z}^{2} & \partial
_{z}^{2}a^{2}\partial _{z}+\partial _{z}a^{2}\partial _{z}^{2}+\partial
_{z}a^{1}\partial _{z}a^{1}\partial _{z}%
\end{pmatrix}
\label{j}
\end{gather}%
and Hamiltonian densities, respectively, are $h_{2}=a^{3}$, $h_{1}=-\frac{1}{%
2}a^{1}(\partial _{z}^{-1}a^{2})^{2}-(\partial _{z}^{-1}a^{2})(\partial
_{z}^{-1}a^{3})$, where $H_{i}=\int h_{i}dz$. These two Hamiltonian
operators $\hat{A}_{1}$ and $\hat{A}_{2}$ are homogeneous (see \cite{DN,DN2}
for more details).

In this paper we introduce new integrable hierarchy, which contains the
above three-component system as a so called \textquotedblleft high
frequency\textquotedblright\ limit. We prove that this intermediate\
dispersive system also is bi-Hamiltonian. In this paper we use the following

\textbf{Theorem} (see detail in \cite{MaksTsar}): \textit{If hydrodynamic
type system written in a conservative form}%
\begin{equation}
u_{t}^{k}=(f^{k}(\mathbf{u}))_{x}  \label{h}
\end{equation}%
\textit{also possesses an extra conservation law} $\partial _{t}p(\mathbf{u}%
)=\partial _{x}g(\mathbf{u})$ \textit{such that (here} $g_{km}$ \textit{is a
constant symmetric nondegenerate matrix)}%
\begin{equation}
p=\frac{1}{2}g_{km}u^{k}u^{m},  \label{p}
\end{equation}%
\textit{then} $u^{k}$ \textit{are flat coordinates,} $p(\mathbf{u})$ \textit{%
is a momentum density, and this hydrodynamic type system has a local
Hamiltonian structure}%
\begin{equation}
u_{t}^{i}=g^{ik}\left( \frac{\partial h}{\partial u^{k}}\right) _{x}.
\label{ham}
\end{equation}

\textbf{Proof}: The conservation law $\partial _{t}p(\mathbf{u})=\partial
_{x}g(\mathbf{u})$ leads to%
\begin{equation*}
\frac{\partial p}{\partial u^{i}}\frac{\partial f^{i}}{\partial u^{k}}%
u_{x}^{k}=\frac{\partial g}{\partial u^{k}}u_{x}^{k}.
\end{equation*}%
Then (see (\ref{p}))%
\begin{equation*}
\frac{\partial g}{\partial u^{k}}=g_{sm}u^{m}\frac{\partial f^{s}}{\partial
u^{k}}=\frac{\partial }{\partial u^{k}}(g_{sm}u^{m}f^{s})-g_{km}f^{m}.
\end{equation*}%
This means that the last term $g_{km}f^{m}$ must be a partial derivative of
some function $h(\mathbf{u})$. Then indeed $f^{i}=g^{ik}\frac{\partial h}{%
\partial u^{k}}$ (cf. (\ref{h}) and (\ref{ham})).

This theorem can be expand to dispersive integrable systems. \textit{If a
conservative dispersive system} (cf. (\ref{h}))%
\begin{equation}
u_{t}^{k}=(f^{k}(\mathbf{u},\mathbf{u}_{x},\mathbf{u}_{xx},...))_{x}
\label{disp}
\end{equation}%
\textit{also has an extra conservation law} $\partial _{t}p(\mathbf{u}%
)=\partial _{x}g(\mathbf{u},\mathbf{u}_{x},\mathbf{u}_{xx},...)$ \textit{%
such that} $p(\mathbf{u})$ \textit{is determined by} (\ref{p}), \textit{then}
(\ref{disp}) \textit{has a local Hamiltonian structure}. Indeed (see the
above Theorem)%
\begin{equation*}
\partial _{x}g(\mathbf{u},\mathbf{u}_{x},\mathbf{u}_{xx},...)=\frac{\partial
p}{\partial u^{k}}(f^{k}(\mathbf{u},\mathbf{u}_{x},\mathbf{u}%
_{xx},...))_{x}=g_{km}u^{m}(f^{k}(\mathbf{u},\mathbf{u}_{x},\mathbf{u}%
_{xx},...))_{x}
\end{equation*}%
\begin{equation*}
=(g_{km}u^{m}f^{k}(\mathbf{u},\mathbf{u}_{x},\mathbf{u}%
_{xx},...))_{x}-g_{km}f^{k}(\mathbf{u},\mathbf{u}_{x},\mathbf{u}%
_{xx},...)u_{x}^{m}.
\end{equation*}%
Now we utilize a well-known relationship (see, for instance, \cite{olver})%
\begin{equation*}
\frac{\delta \mathbf{H}}{\delta u^{k}}u_{x}^{k}=(Q(\mathbf{u},\mathbf{u}_{x},%
\mathbf{u}_{xx},...))_{x}.
\end{equation*}%
Thus a function $h(\mathbf{u},\mathbf{u}_{x},\mathbf{u}_{xx},...)$ exists if%
\begin{equation*}
g_{km}f^{m}(\mathbf{u},\mathbf{u}_{x},\mathbf{u}_{xx},...)=\frac{\delta 
\mathbf{H}}{\delta u^{k}},
\end{equation*}%
where $\mathbf{H}=\int h(\mathbf{u},\mathbf{u}_{x},\mathbf{u}_{xx},...)dx$.
This means that dispersive system (\ref{disp}) takes the Hamiltonian form%
\begin{equation*}
u_{t}^{k}=g^{km}\left( \frac{\delta \mathbf{H}}{\delta u^{m}}\right) _{x}.
\end{equation*}

The operator $\hat{A}_{1}$ was found in \cite{FM}, and it is completely
specified by its leading term, which is a contravariant flat
pseudo-Riemannian metric $g^{ik}$. The observation that led to finding $\hat{%
A}_{1}$ was that the eigenvalues $u^{k}(\mathbf{a})$ of the matrix $\mathbf{B%
}$ are conservation law densities in the Lax pair of the system (\ref{wdvv})%
\begin{equation}
\mathbf{\psi }_{z}=\lambda \mathbf{B\psi }=\lambda 
\begin{pmatrix}
0 & 1 & 0 \\ 
a^{2} & a^{1} & 1 \\ 
a^{3} & a^{2} & 0%
\end{pmatrix}%
\mathbf{\psi },\qquad \mathbf{\psi }_{y}=\lambda \mathbf{C\psi }=\lambda 
\begin{pmatrix}
0 & 0 & 1 \\ 
a^{3} & a^{2} & 0 \\ 
(a^{2})^{2}-a^{1}a^{3} & a^{3} & 0%
\end{pmatrix}%
\mathbf{\psi .}  \label{a}
\end{equation}%
If the system is rewritten using the above eigenvalues as new dependent
variables $u^{k}$, i.e., using the point transformation%
\begin{equation}
a^{1}=u^{1}+u^{2}+u^{3},\text{ \ }a^{2}=-\frac{1}{2}%
(u^{1}u^{2}+u^{1}u^{3}+u^{2}u^{3}),\text{ \ }a^{3}=u^{1}u^{2}u^{3},
\label{viet}
\end{equation}%
the operator $\hat{A}_{1}$ becomes evident and is of the type $%
A_{1}^{ij}=K^{ij}\partial _{x}^{{}}$, where 
\begin{equation}
K=\frac{1}{2}\left( 
\begin{array}{ccc}
1 & -1 & -1 \\ 
-1 & 1 & -1 \\ 
-1 & -1 & 1%
\end{array}%
\right) ,  \label{key}
\end{equation}%
and the Hamiltonian density is $h_{2}=u^{1}u^{2}u^{3}$. In these new
coordinates WDVV\ associativity system~(\ref{wdvv}) takes the form%
\begin{equation}
u_{y}^{1}=\frac{1}{2}(u^{2}u^{3}-u^{1}u^{2}-u^{1}u^{3})_{z},\text{ \ }%
u_{y}^{2}=\frac{1}{2}(u^{1}u^{3}-u^{1}u^{2}-u^{2}u^{3})_{z},\text{ \ }%
u_{y}^{3}=\frac{1}{2}(u^{1}u^{2}-u^{1}u^{3}-u^{2}u^{3})_{z}.  \label{ha}
\end{equation}

The quadratic relationship (see (\ref{viet}))%
\begin{equation}
a^{2}=-\frac{1}{2}(u^{1}u^{2}+u^{1}u^{3}+u^{2}u^{3})  \label{kvadro}
\end{equation}%
is nothing but (\ref{p}) for the WDVV associativity system (\ref{wdvv}),
where $g^{ij}=K^{ij}$.

\section{Transformation to the Dispersionful Version}

\label{sec:trans}

In this Section we consider\ the left matrix equation of linear problem (\ref%
{a})%
\begin{equation*}
\mathbf{\psi }_{z}=\lambda 
\begin{pmatrix}
0 & 1 & 0 \\ 
a^{2} & a^{1} & 1 \\ 
a^{3} & a^{2} & 0%
\end{pmatrix}%
\mathbf{\psi }.
\end{equation*}%
Instead of three linear equations%
\begin{equation*}
\psi _{z}=\lambda \psi _{1},\text{ \ }\psi _{1,z}=\lambda a^{2}\psi +\lambda
a^{1}\psi _{1}+\lambda \psi _{2},\text{ \ }\psi _{2,z}=\lambda a^{3}\psi
+\lambda a^{2}\psi _{1},
\end{equation*}%
we shall use a single scalar ordinary equation of third order%
\begin{equation}
\psi _{zzz}=\lambda a^{1}\psi _{zz}+(2\lambda ^{2}a^{2}+\lambda
a_{z}^{1})\psi _{z}+(\lambda ^{2}a_{z}^{2}+\lambda ^{3}a^{3})\psi ,
\label{tri}
\end{equation}%
where $\psi _{1}=\lambda ^{-1}\psi _{z},\psi _{2}=\lambda ^{-2}\psi
_{zz}-\lambda ^{-1}a^{1}\psi _{z}-a^{2}\psi $.

Now we introduce the transformation%
\begin{equation}
z=\epsilon (e^{x/\epsilon }-1),  \label{z}
\end{equation}%
where $\epsilon $ is an arbitrary parameter. Then $\partial _{z}\rightarrow
e^{-x/\epsilon }\partial _{x}$. If $\epsilon \rightarrow \infty $, then $%
z(x,\epsilon )\rightarrow x$. In this case (\ref{tri}) takes the form%
\begin{equation}
\psi _{xxx}-\left( \frac{3}{\epsilon }+\lambda w^{1}\right) \psi _{xx}
\label{tre}
\end{equation}%
\begin{equation*}
+\left[ \frac{2}{\epsilon ^{2}}+\lambda \left( \frac{2}{\epsilon }%
w^{1}-w_{x}^{1}\right) -2\lambda ^{2}w^{2}\right] \psi _{x}+\left[ \lambda
^{2}\left( \frac{2}{\epsilon }w^{2}-w_{x}^{2}\right) -\lambda ^{3}w^{3}%
\right] \psi =0,
\end{equation*}%
where we introduced new field variables $w^{k}$ and $v^{m}$ instead of $%
a^{k} $ and $u^{m}$, respectively (see (\ref{wdvv}), (\ref{a}), (\ref{ha})):%
\begin{equation*}
a^{1}=w^{1}e^{-x/\epsilon },\text{ \ }a^{2}=w^{2}e^{-2x/\epsilon },\text{ \ }%
a^{3}=w^{3}e^{-3x/\epsilon },\text{ \ }u^{k}=v^{k}e^{-x/\epsilon },\text{ }%
k=1,2,3.
\end{equation*}%
So variable coefficients of third order differential equation (\ref{tre}) do
not depend explicitly on independent variable \textquotedblleft $x$%
\textquotedblright\ as well as in (\ref{tri}). This equation (\ref{tre})
also can be written as a linear problem (\ref{a}) in the matrix form%
\begin{equation}
\mathbf{\tilde{\psi}}_{x}=\left[ \lambda 
\begin{pmatrix}
0 & 1 & 0 \\ 
w^{2} & w^{1} & 1 \\ 
w^{3} & w^{2} & 0%
\end{pmatrix}%
+\frac{1}{\epsilon }%
\begin{pmatrix}
0 & 0 & 0 \\ 
0 & 1 & 0 \\ 
0 & 0 & 2%
\end{pmatrix}%
\right] \mathbf{\tilde{\psi}},  \label{b}
\end{equation}%
where (the column of the vector function $\mathbf{\tilde{\psi}}$ contains
three components $\tilde{\psi}=\psi ,\tilde{\psi}_{1}$ and $\tilde{\psi}_{2}$%
, consequently)%
\begin{equation*}
\tilde{\psi}_{1}=\frac{1}{\lambda }\psi _{x},\text{ \ \ }\tilde{\psi}_{2}=%
\frac{1}{\lambda ^{2}}\psi _{xx}-\frac{1}{\lambda }\left( \frac{1}{\epsilon
\lambda }+w^{1}\right) \psi _{x}-w^{2}\psi .
\end{equation*}

Below in this Section we consider construction of an integrable hierarchy
dependent on the extra parameter $\epsilon $.

\section{The Intermediate\ Integrable Hierarchy}

\label{sec:inter}

In comparison with a standard construction of integrable systems based on a
compatibility condition of a corresponding Lax pair, in this Section we
follow to another strategy:

1. We compute first conservation law densities from a generating function of
conservation law densities;

2. We select quadratic relationships of type (\ref{p});

3. We consider the conservation law densities involved in these quadratic
relationships as flat coordinates and momenta, we can take any other
conservation law densities as Hamiltonian densities to create corresponding
Hamiltonian systems;

4. A compatible pair of Hamiltonian operators must determine each member of
an integrable hierarchy by an appropriate choice of corresponding pairs of
Hamiltonian densities. We present such a pair of Hamiltonian densities for
the intermediate dispersive system.

\subsection{A Generating Function of Conservation Law Densities}

\label{subsec:density}

The substitution%
\begin{equation}
\psi =e^{\int rdx}  \label{r}
\end{equation}%
into third order linear differential equation (\ref{tre}) leads to the
second order ordinary nonlinear differential equation%
\begin{equation}
r_{xx}+3rr_{x}+r^{3}-\left( \frac{3}{\epsilon }+\lambda w^{1}\right)
(r_{x}+r^{2})  \label{ar}
\end{equation}%
\begin{equation*}
+\left[ \frac{2}{\epsilon ^{2}}+\lambda \left( \frac{2}{\epsilon }%
w^{1}-w_{x}^{1}\right) -2\lambda ^{2}w^{2}\right] r+\lambda ^{2}\left( \frac{%
2}{\epsilon }w^{2}-w_{x}^{2}-\lambda w^{3}\right) =0,
\end{equation*}%
where the function $r$ is a generating function of conservation law
densities. Indeed, quasipolynomial conservation law densities can be
obtained by substitution of the Laurent series ($\lambda \rightarrow \infty $%
)%
\begin{equation}
r=\lambda a_{-1}+a_{0}+\frac{a_{1}}{\lambda }+\frac{a_{2}}{\lambda }+\frac{%
a_{3}}{\lambda ^{3}}+...,  \label{branch}
\end{equation}%
where all coefficients $a_{k}(\mathbf{w},\mathbf{w}_{x},\mathbf{w}_{xx},...)$
can be found iteratively:%
\begin{equation}
(a_{-1})^{3}-(a_{-1})^{2}w^{1}-2a_{-1}w^{2}-w^{3}=0,  \label{first}
\end{equation}%
\begin{equation}
\lbrack 3(a_{-1})^{2}-2a_{-1}w^{1}-2w^{2}]\left( \frac{1}{\epsilon }%
-a_{0}\right) =(3a_{-1}-w^{1}){a_{-1,x}}-a_{-1}w_{x}^{1}-w_{x}^{2},
\label{sec}
\end{equation}%
\begin{equation}
\lbrack 3(a_{-1})^{2}-2a_{-1}w^{1}-2w^{2}]a_{1}+\left( \frac{2}{\epsilon ^{2}%
}-\frac{6}{\epsilon }a_{0}+3(a_{0})^{2}\right) a_{-1}+\left( \frac{2}{%
\epsilon }-a_{0}\right) a_{0}w^{1}  \label{third}
\end{equation}%
\begin{equation*}
+(3a_{-1}-w^{1}){a_{0,x}}+3\left( a_{0}-\frac{1}{\epsilon }\right) {a_{-1,x}}%
-a_{0}w_{x}^{1}+{a_{-1,xx}}=0,...
\end{equation*}

The first equation (\ref{first}) determines three roots $v^{m}$, if one
assumes that functions $w^{k}$ are given. Or we can consider these roots $%
v^{m}$ as given functions, then functions $w^{k}$ are polynomials according
to the Vi\`{e}te theorem, i.e. (cf. (\ref{viet}))%
\begin{equation}
w^{1}=v^{1}+v^{2}+v^{3},\text{ \ }w^{2}=-\frac{1}{2}%
(v^{1}v^{2}+v^{1}v^{3}+v^{2}v^{3}),\text{ \ }w^{3}=v^{1}v^{2}v^{3}.
\label{vieta}
\end{equation}%
Existence of these three roots $v^{m}$ means existence of three branches of
conservation law densities (see (\ref{branch}))%
\begin{equation}
r^{(k)}=\lambda v^{k}+a_{0}^{(k)}+\frac{a_{1}^{(k)}}{\lambda }+\frac{%
a_{2}^{(k)}}{\lambda }+\frac{a_{3}^{(k)}}{\lambda ^{3}}+...,\text{ \ }%
k=1,2,3,  \label{ra}
\end{equation}%
where all other higher conservation law densities $a_{k}^{(m)}$ can be
expressed as rational functions with respect to three roots $v^{k}$ and
their higher derivatives. For instance (see (\ref{sec}) and (\ref{third}),
respectively)%
\begin{equation*}
a_{0}^{(1)}=\frac{1}{\epsilon }-\frac{1}{2}\partial _{x}\ln
[(v^{1}-v^{2})(v^{1}-v^{3})],
\end{equation*}%
\begin{equation*}
a_{1}^{(1)}=\frac{(\frac{2}{\epsilon }-\partial _{x})\left[
(2v^{1}-v^{2}-v^{3})a_{0}^{(1)}+(\partial _{x}-\frac{1}{\epsilon })v^{1}%
\right] -(2v^{1}-v^{2}-v^{3})\left( a_{0}^{(1)}\right) ^{2}}{%
(v^{1}-v^{2})(v^{1}-v^{3})}
\end{equation*}%
\begin{equation*}
a_{2}^{(1)}=\frac{(\partial _{x}-\frac{2}{\epsilon })\left[ \frac{3}{2}%
\left( a_{0}^{(1)}\right) ^{2}+(\partial _{x}-\frac{1}{\epsilon }%
)a_{0}^{(1)}+(2v^{1}-v^{2}-v^{3})a_{1}^{(1)}\right] +\left(
a_{0}^{(1)}\right) ^{3}+2(2v^{1}-v^{2}-v^{3})a_{0}^{(1)}a_{1}^{(1)}}{%
(v^{1}-v^{2})(v^{1}-v^{3})}.
\end{equation*}

Lower (nonlocal) conservation law densities can be found by substitution
another expansion ($\lambda \rightarrow 0$)%
\begin{equation}
r=\lambda ^{2}(q_{0}+\lambda q_{1}+\lambda ^{2}q_{2}+\lambda ^{3}q_{3}+...),
\label{ro}
\end{equation}%
into (\ref{ar}). Corresponding conservation law densities $q_{k}$ cannot be
expressed via roots $v^{k}$ and their finite number of derivatives, i.e.%
\begin{equation}
\left( \partial _{x}-\frac{1}{\epsilon }\right) q_{0}=w^{2},  \label{raz}
\end{equation}%
\begin{equation}
\left( \partial _{x}-\frac{1}{\epsilon }\right) \left( \partial _{x}-\frac{2%
}{\epsilon }\right) q_{1}=w^{3}-\frac{2}{\epsilon }%
q_{0}w^{1}+(q_{0}w^{1})_{x},  \label{dva}
\end{equation}%
\begin{equation}
\left( \partial _{x}-\frac{1}{\epsilon }\right) \left( \partial _{x}-\frac{2%
}{\epsilon }\right) q_{2}=\frac{3}{\epsilon }(q_{0})^{2}-\frac{2}{\epsilon }%
q_{1}w^{1}+2q_{0}w^{2}+\left( q_{1}w^{1}-\frac{3}{2}(q_{0})^{2}\right) _{x},
\label{kva}
\end{equation}%
\begin{equation}
\left( \partial _{x}-\frac{1}{\epsilon }\right) \left( \partial _{x}-\frac{2%
}{\epsilon }\right) q_{3}=(q_{0})^{2}w^{1}+2q_{1}w^{2}+\left( \partial _{x}-%
\frac{2}{\epsilon }\right) (q_{2}w^{1})-3\left( \partial _{x}-\frac{2}{%
\epsilon }\right) (q_{1}q_{0}),...  \label{kwa}
\end{equation}

However some important observations one can made analyzing this expansion
and Vi\`{e}te's formulas (\ref{vieta}). Indeed:

1. $w^{1}$ is a conservation law density, because $w^{1}$ is a linear
combination of roots $v^{k}$, which are conservation law density according
to the construction (i.e. they are first elements of expansions (\ref{ra})
of the generation function of conservation law densities $r$);

2. the first relationship (\ref{raz}) means that $w^{2}$ is a conservation
law density, because $q_{0}$ is a conservation law density and any
conservation law density is determined up to total derivatives;

3. thus the central quadratic relationship in (\ref{vieta}) means that any
integrable system associated with (\ref{tre}) or (\ref{b}) can possess a 
\textit{first} local Hamiltonian structure (see (\ref{p}) and (\ref{ham}));

4. the second relationship (\ref{dva}) means that the expression $w^{3}-%
\frac{2}{\epsilon }q_{0}w^{1}$ is a conservation law density, because $q_{1}$
is a conservation law density and any conservation law density is determined
up to total derivatives;

5. the third relationship (\ref{kva}) means that any integrable system
associated with (\ref{tre}) can possess a \textit{second} local Hamiltonian
structure, because (see (\ref{p}) and (\ref{ham})) $q_{2}$ is a quadratic
expression in terms of functions $w^{1},q_{0},q_{1}$ (all these four
functions $w^{1},q_{0},q_{1},q_{2}$ are conservation law densities) up to a
total derivative:%
\begin{equation}
-\frac{1}{\epsilon }q_{2}+(q_{2})_{x}=q_{1}w^{1}-\frac{1}{2}(q_{0})^{2},
\label{quadro}
\end{equation}%
Here we took into account that the conservation law density $w^{2}$ can be
expressed via $q_{0}$ from (\ref{raz}).

\subsection{A Non-Evolutionary Form}

\label{subsec:evolu}

In this Subsection we are going to construct an integrable system whose the
so called high-frequency limit ($\epsilon \rightarrow \infty $) is precisely
original WDVV associativity system (\ref{wdvv}). In such a case, this
dispersive integrable system can be written, for instance, via flat
coordinates $v^{k}$. Its Hamiltonian density is $w^{3}-\frac{2}{\epsilon }%
q_{0}w^{1}$, because the Hamiltonian density of (\ref{wdvv}) is $w^{3}$ (see 
\cite{FGMN}). However, taking into account that (see (\ref{raz})) the
relationship between field variables $q_{0}$ and $w^{2}$ is uninvertible in
a local sense, the Hamiltonian density becomes nonlocal:%
\begin{equation}
h_{2}=v^{1}v^{2}v^{3}+\frac{1}{\epsilon }(v^{1}+v^{2}+v^{3})\left( \partial
_{x}-\frac{1}{\epsilon }\right) ^{-1}(v^{1}v^{2}+v^{1}v^{3}+v^{2}v^{3}).
\label{hamilt}
\end{equation}%
Expanding in $\epsilon $, this Hamiltonian density reads ($\epsilon
\rightarrow \infty $)%
\begin{equation*}
h_{2}=v^{1}v^{2}v^{3}+\frac{1}{\epsilon }(v^{1}+v^{2}+v^{3})\partial
_{x}^{-1}\left( 1+\frac{1}{\epsilon }\partial _{x}^{-1}+\frac{1}{\epsilon
^{2}}\partial _{x}^{-2}+\frac{1}{\epsilon ^{3}}\partial _{x}^{-3}+...\right)
(v^{1}v^{2}+v^{1}v^{3}+v^{2}v^{3}).
\end{equation*}%
Under the differential substitutions%
\begin{equation}
v^{k}=\left( \partial _{x}+\frac{1}{\epsilon }\right) \eta ^{k},  \label{vik}
\end{equation}%
Hamiltonian density (\ref{hamilt}) reduces to the local form%
\begin{equation*}
h_{2}=v^{1}v^{2}v^{3}-\frac{1}{\epsilon }(v^{1}v^{2}+v^{1}v^{3}+v^{2}v^{3})(%
\eta ^{1}+\eta ^{2}+\eta ^{3}),
\end{equation*}%
while local Hamiltonian structure (see (\ref{key})) becomes nonlocal:%
\begin{equation*}
\eta _{t}^{i}=-K^{im}\left( \partial _{x}^{2}-\frac{1}{\epsilon ^{2}}\right)
^{-1}\left( \frac{\delta \mathbf{H}_{2}}{\delta \eta ^{m}}\right) _{x}.
\end{equation*}%
This means that the intermediate\ dispersive system has a non-evolutionary
form%
\begin{equation*}
\frac{1}{\epsilon ^{2}}\eta _{t}^{i}-\eta _{xxt}^{i}=K^{im}\left( \frac{%
\delta \mathbf{H}_{2}}{\delta \eta ^{m}}\right) _{x}.
\end{equation*}

\textbf{Remark}: The intermediate\ dispersive system also has a local
Lagrangian representation%
\begin{equation}
S_{2}=\int \left[ \frac{1}{2}K_{im}\left( \frac{1}{\epsilon ^{2}}\phi
_{x}^{i}-\phi _{xxx}^{i}\right) \phi _{t}^{m}-h_{2}\right] dxdt,  \label{k}
\end{equation}%
where $\phi ^{k}$ are potentials such that $\eta ^{k}=\phi _{x}^{k}$ and $%
K_{ij}$ is determined by (\ref{key}).

Also independently taking into account (\ref{quadro}) the intermediate\
dispersive system can be written in the second Hamiltonian form%
\begin{equation*}
w_{t}^{1}=\left( \frac{\delta \mathbf{H}_{1}}{\delta q_{1}}\right) _{x},%
\text{ \ \ }q_{0,t}=-\left( \frac{\delta \mathbf{H}_{1}}{\delta q_{0}}%
\right) _{x},\text{ \ \ }q_{1,t}=\left( \frac{\delta \mathbf{H}_{1}}{\delta
w^{1}}\right) _{x},
\end{equation*}%
where (see (\ref{kwa}))%
\begin{equation*}
\mathbf{H}_{1}=\int \left[ \frac{1}{2}(q_{0})^{2}\eta _{x}+q_{1}q_{0,x}+%
\frac{1}{\epsilon }q_{1}\left( 2q_{0}+\eta \eta _{x}+\frac{1}{\epsilon }\eta
^{2}\right) \right] dx,
\end{equation*}%
where we introduced the field variable $\eta $ such that (see (\ref{vik})
and the first relationship in (\ref{vieta}))%
\begin{equation*}
w^{1}=\left( \partial _{x}+\frac{1}{\epsilon }\right) \eta \text{ \ \ }%
\leftrightarrow \text{ \ \ }\eta =\eta ^{1}+\eta ^{2}+\eta ^{3}.
\end{equation*}%
In field variables $q_{0},q_{1},\eta $ the intermediate\ dispersive system
takes the non-evolutionary form%
\begin{equation}
\left( \partial _{x}+\frac{1}{\epsilon }\right) \eta _{t}=\left( q_{0,x}+%
\frac{1}{\epsilon }(2q_{0}+\eta \eta _{x})+\frac{1}{\epsilon ^{2}}\eta
^{2}\right) _{x},\text{ \ \ }q_{0,t}=\left( q_{1,x}-q_{0}\eta _{x}-\frac{2}{%
\epsilon }q_{1}\right) _{x},  \label{non}
\end{equation}%
\begin{equation*}
\left( \partial _{x}-\frac{1}{\epsilon }\right) q_{1,t}=\left( \frac{1}{2}%
[(q_{0})^{2}]_{x}+\frac{1}{\epsilon }\eta q_{1,x}-\frac{2}{\epsilon ^{2}}%
\eta q_{1}\right) _{x},
\end{equation*}%
where above Hamiltonian structure becomes nonlocal:%
\begin{equation*}
\eta _{t}=\left( \partial _{x}+\frac{1}{\epsilon }\right) ^{-1}\left( \frac{%
\delta \mathbf{H}_{1}}{\delta q_{1}}\right) _{x},\text{ \ \ }q_{0,t}=-\left( 
\frac{\delta \mathbf{H}_{1}}{\delta q_{0}}\right) _{x},\text{ \ \ }%
q_{1,t}=-\left( \partial _{x}-\frac{1}{\epsilon }\right) ^{-1}\left( \frac{%
\delta \mathbf{H}_{1}}{\delta \eta }\right) _{x}.
\end{equation*}%
Under the potential substitutions%
\begin{equation*}
\eta =\left( \partial _{x}-\frac{1}{\epsilon }\right) \zeta _{x},\text{ \ }%
q_{0}=Q_{0,x},\text{ \ }q_{1}=Q_{1,x}
\end{equation*}%
we obtain another local Lagrangian representation%
\begin{equation}
S_{1}=\int \left[ \frac{1}{2}\left( \zeta _{xxx}-\frac{1}{\epsilon ^{2}}%
\zeta _{x}\right) Q_{1,t}+\frac{1}{2}\left( Q_{1,xxx}-\frac{1}{\epsilon ^{2}}%
Q_{1,x}\right) \zeta _{t}-\frac{1}{2}Q_{0,x}Q_{0,t}-h_{1}\right] dxdt.
\label{s}
\end{equation}

\subsection{The WDVV Associativity System as a High-Frequency Limit}

\label{subsec:wdvv}

Higher and lower conservation law densities for the WDVV associativity
system (\ref{wdvv}) can be obtained (see transformation (\ref{z})) in the 
\textit{high-frequency limit} $\epsilon \rightarrow \infty $. This means
that expansion (\ref{ra}) leads to little bit more simple differential
consequences:%
\begin{equation*}
a_{0}^{(1)}=-\frac{1}{2}(\ln (u^{1}-u^{2})(u^{1}-u^{3}))_{z},
\end{equation*}%
\begin{equation*}
a_{1}^{(1)}=-\frac{\left[ (2u^{1}-u^{2}-u^{3})a_{0}^{(1)}+u_{z}^{1}\right]
_{z}+(2u^{1}-u^{2}-u^{3})\left( a_{0}^{(1)}\right) ^{2}}{%
(u^{1}-u^{2})(u^{1}-u^{3})},
\end{equation*}%
\begin{equation*}
a_{2}^{(1)}=\frac{\left[ \frac{3}{2}\left( a_{0}^{(1)}\right) ^{2}+\left(
a_{0}^{(1)}\right) _{z}+(2u^{1}-u^{2}-u^{3})a_{1}^{(1)}\right] _{z}+\left(
a_{0}^{(1)}\right) ^{3}+2(2u^{1}-u^{2}-u^{3})a_{0}^{(1)}a_{1}^{(1)}}{%
(u^{1}-u^{2})(u^{1}-u^{3})},...
\end{equation*}%
where $u^{k}$ are roots (see (\ref{vieta})) of third order algebraic
equation (\ref{first}). They are nothing but flat coordinates of first
Hamiltonian structure (see (\ref{i}), (\ref{viet}), (\ref{key}), (\ref{ha}))
of WDVV associativity system (\ref{wdvv}). Coefficients of expansion (\ref%
{ro}) significantly simplify (cf. (\ref{raz}), (\ref{dva}), (\ref{kva})):%
\begin{equation}
q_{0,z}=w^{2},  \label{eins}
\end{equation}%
\begin{equation}
q_{1,zz}=w^{3}+(q_{0}w^{1})_{z},  \label{cwai}
\end{equation}%
\begin{equation}
q_{2,zz}=2q_{0}w^{2}+\left( q_{1}w^{1}-\frac{3}{2}(q_{0})^{2}\right) _{z},
\label{drai}
\end{equation}%
\begin{equation}
q_{3,zz}=(q_{2}w^{1}-3q_{1}q_{0})_{z}+(q_{0})^{2}w^{1}+2q_{1}w^{2},...
\label{fix}
\end{equation}%
Taking into account that $w^{2}=q_{0,z}$, equation (\ref{drai}) can be
integrated once:%
\begin{equation}
q_{2,z}=q_{1}w^{1}-\frac{1}{2}(q_{0})^{2}.  \label{fur}
\end{equation}%
WDVV associativity system (\ref{wdvv}) takes the form%
\begin{equation}
w_{y}^{1}=q_{0,zz},\text{ \ \ }q_{0,y}=(q_{1,z}-q_{0}w^{1})_{z},\text{ \ \ }%
q_{1,y}=\frac{1}{2}[(q_{0})^{2}]_{z},  \label{wdvv2}
\end{equation}%
where we took into account that $w^{3}=(q_{1,z}-q_{0}w^{1})_{z}$ (see (\ref%
{cwai})).

Since $w^{1},q_{0},q_{1}$ are conservation law densities, quadratic
relationship (\ref{fur}) ($q_{2,z}$ is a trivial conservation law density)
again means that the high-frequency limit (i.e. WDVV associativity system
and all its commuting flows) possesses a bi-Hamiltonian structure, as it was
earlier proved in \cite{FGMN}). Thus evolutionary system (\ref{wdvv2}) can
be written in the Hamiltonian form%
\begin{equation}
w_{y}^{1}=\left( \frac{\delta \mathbf{H}_{1}}{\delta q_{1}}\right) _{z},%
\text{ \ \ }q_{0,y}=-\left( \frac{\delta \mathbf{H}_{1}}{\delta q_{0}}%
\right) _{z},\text{ \ \ }q_{1,y}=\left( \frac{\delta \mathbf{H}_{1}}{\delta
w^{1}}\right) _{z},  \label{locham}
\end{equation}%
where the Hamiltonian density $h_{1}=q_{1}q_{0,z}+\frac{1}{2}%
(q_{0})^{2}w^{1} $ (see (\ref{fix}), here we substituted $w^{2}$ from (\ref%
{eins})). Under differential substitutions (\ref{eins}), (\ref{cwai}) in the
original field variables $w^{k}$ Hamiltonian structure (\ref{locham}) of
evolutionary system (\ref{wdvv2}) becomes precisely (\ref{j}). This confirms
that WDVV associativity system (\ref{wdvv}) possesses a bi-Hamiltonian
structure.

\section{A Dispersionless Limit of Intermediate\ Dispersive System}

\label{sec:expand}

To derive a correct dispersionless limit ($\epsilon \rightarrow 0$) first we
need to re-scale a spectral parameter $\lambda \rightarrow \Lambda /\epsilon 
$. Then substitution (cf. (\ref{r}))

\begin{equation*}
\psi =\exp \left( \frac{1}{\epsilon }\int rdx\right)
\end{equation*}%
into (\ref{tre}) leads to the second order ordinary nonlinear differential
equation (cf. (\ref{ar}))%
\begin{equation}
\epsilon ^{2}r_{xx}+3\epsilon rr_{x}+r^{3}-\left( 3+\Lambda w^{1}\right)
(\epsilon r_{x}+r^{2})  \label{quasi}
\end{equation}%
\begin{equation*}
+\left[ 2+\Lambda \left( 2w^{1}-\epsilon w_{x}^{1}\right) -2\Lambda ^{2}w^{2}%
\right] r+\Lambda ^{2}\left( 2w^{2}-\epsilon w_{x}^{2}-\Lambda w^{3}\right)
=0.
\end{equation*}%
Then the first expansion (cf. (\ref{branch}))%
\begin{equation}
r=\Lambda b_{-1}+b_{0}+\frac{b_{1}}{\Lambda }+\frac{b_{2}}{\Lambda ^{2}}+%
\frac{b_{3}}{\Lambda ^{3}}+...,\text{ \ \ }\Lambda \rightarrow \infty
\label{ras}
\end{equation}%
yields infinitely many differential consequences, which slightly different
in comparison with (\ref{first}), (\ref{sec}), (\ref{third}):%
\begin{equation*}
(b_{-1})^{3}-(b_{-1})^{2}w^{1}-2b_{-1}w^{2}-w^{3}=0,
\end{equation*}%
\begin{equation*}
\lbrack 3(b_{-1})^{2}-2b_{-1}w^{1}-2w^{2}](1-b_{0})=\epsilon \lbrack
(3b_{-1}-w^{1}){b_{-1,x}}-b_{-1}w_{x}^{1}-w_{x}^{2}],
\end{equation*}%
\begin{equation*}
\lbrack
3(b_{-1})^{2}-2b_{-1}w^{1}-2w^{2}]b_{1}+[2-6b_{0}+3(b_{0})^{2}]b_{-1}+(2-b_{0})b_{0}w^{1}
\end{equation*}%
\begin{equation*}
+\epsilon \lbrack (3b_{-1}-w^{1}){b_{0,x}}+3(b_{0}-1){b_{-1,x}}%
-b_{0}w_{x}^{1}]+\epsilon ^{2}b{_{-1,xx}}=0,...
\end{equation*}%
The second expansion (cf. (\ref{ro}))%
\begin{equation}
r=\Lambda ^{2}(c_{0}+\Lambda c_{1}+\Lambda ^{2}c_{2}+\Lambda ^{3}c_{3}+...),%
\text{ \ \ }\Lambda \rightarrow 0  \label{dwa}
\end{equation}%
yields (cf. (\ref{raz}), (\ref{dva}), (\ref{kva}), (\ref{kwa})):%
\begin{equation}
(1-\epsilon \partial _{x})c_{0}=-w^{2},  \label{uno}
\end{equation}%
\begin{equation}
(1-\epsilon \partial _{x})(2-\epsilon \partial
_{x})c_{1}=w^{3}-2c_{0}w^{1}+\epsilon (c_{0}w^{1})_{x},  \label{due}
\end{equation}%
\begin{equation}
(1-\epsilon \partial _{x})(2-\epsilon \partial
_{x})c_{2}=3(c_{0})^{2}-2c_{1}w^{1}+2c_{0}w^{2}+\epsilon \left( c_{1}w^{1}-%
\frac{3}{2}(c_{0})^{2}\right) _{x},  \label{trua}
\end{equation}%
\begin{equation}
(1-\epsilon \partial _{x})(2-\epsilon \partial
_{x})c_{3}=(c_{0})^{2}w^{1}+2c_{1}w^{2}+(2-\epsilon \partial
_{x})(3c_{0}c_{1}-c_{2}w^{1}),...  \label{quatro}
\end{equation}

\subsection{A Egorov bi-Hamiltonian Hydrodynamic Type System}

\label{subsec:disperless}

In the dispersionless limit, the algebraic relationship%
\begin{equation}
r^{3}-\left( 3+\Lambda w^{1}\right) r^{2}+2(1+\Lambda w^{1}-\Lambda
^{2}w^{2})r+2\Lambda ^{2}w^{2}-\Lambda ^{3}w^{3}=0  \label{ellip}
\end{equation}%
follows from (\ref{quasi}). Infinitely many \textit{rational} conservation
law densities can be obtained by substitution (\ref{ras}) into (\ref{ellip}):%
\begin{equation*}
b_{0}=1,\text{ \ \ }b_{1}^{(1)}=-\frac{v^{2}+v^{3}}{%
(v^{1}-v^{2})(v^{1}-v^{3})},\text{ \ \ }b_{2}^{(1)}=0,
\end{equation*}%
\begin{equation*}
b_{3}^{(1)}=\frac{({v^{2}}+{v^{3}})\left( ({v^{1}})^{2}-{v^{1}}({v^{2}}+{%
v^{3}})+({v^{2}})^{2}+({v^{3}})^{2}+{v^{2}v^{3}}\right) }{({v^{1}}-{v^{2}}%
)^{3}({v^{1}}-{v^{3}})^{3}},...
\end{equation*}%
where $v^{k}$ are roots of the cubic algebraic equation (cf. (\ref{first}))%
\begin{equation}
(b_{-1})^{3}-(b_{-1})^{2}w^{1}-2b_{-1}w^{2}-w^{3}=0.  \label{cubic}
\end{equation}%
Infinitely many \textit{polynomial} conservation law densities (cf. with the
high-frequency limit: (\ref{eins}), (\ref{cwai}), (\ref{drai}), (\ref{fix}))
can be obtained by substitution (\ref{dwa}) into (\ref{ellip}):%
\begin{equation}
c_{0}=-w^{2},  \label{one}
\end{equation}%
\begin{equation*}
c_{1}=\frac{1}{2}w^{3}-c_{0}w^{1},
\end{equation*}%
\begin{equation}
c_{2}=\frac{3}{2}(c_{0})^{2}-c_{1}w^{1}+c_{0}w^{2},  \label{two}
\end{equation}%
\begin{equation*}
c_{3}=\frac{1}{2}(c_{0})^{2}w^{1}+c_{1}w^{2}+3c_{0}c_{1}-c_{2}w^{1},...
\end{equation*}%
In this dispersionless limit we again select quadratic relationships (\ref{p}%
) for explicit computation of a bi-Hamiltonian structure.

Since roots $v^{k}$ are conservation law densities, then $%
w^{1}=v^{1}+v^{2}+v^{3}$ is again conservation law density (according to the
Vi\`{e}te theorem, see (\ref{viet}) and (\ref{cubic})). Since $c_{0}$ is a
conservation law density, then also $w^{2}$ is a conservation law density.
Since (according to the Vi\`{e}te theorem, see (\ref{viet}), (\ref{kvadro})
and (\ref{cubic}))%
\begin{equation*}
w^{2}=-\frac{1}{2}(v^{1}v^{2}+v^{1}v^{3}+v^{2}v^{3}),
\end{equation*}%
a dispersionless limit of the intermediate\ system possesses a first local
Hamiltonian structure%
\begin{equation*}
v_{t}^{i}=K^{im}\left( \frac{\partial h_{2}}{\partial v^{m}}\right) _{x},
\end{equation*}%
where the Hamiltonian density $h_{2}=2c_{1}=w^{3}+2w^{1}w^{2}$ (cf. (\ref%
{hamilt})). Thus corresponding hydrodynamic type system takes the symmetric
form%
\begin{equation}
v_{t}^{k}=2\left( \frac{(v^{k})^{2}}{2}-w^{2}\right) _{x},\text{ \ }k=1,2,3.
\label{sys}
\end{equation}

Also (\ref{two}) can be written in the quadratic form (see (\ref{p}))%
\begin{equation*}
c_{2}=\frac{1}{2}(w^{2})^{2}-c_{1}w^{1},
\end{equation*}%
where we took into account (\ref{one}). Thus hydrodynamic type system (\ref%
{sys}) possesses a second local Hamiltonian structure (cf. (\ref{locham}))%
\begin{equation}
w_{t}^{1}=\left( \frac{\partial h_{1}}{\partial c_{1}}\right) _{x},\text{ \
\ }w_{t}^{2}=-\left( \frac{\partial h_{1}}{\partial w^{2}}\right) _{x},\text{
\ \ }c_{1,t}=\left( \frac{\partial h_{1}}{\partial w^{1}}\right) _{x},
\label{egor}
\end{equation}%
where the Hamiltonian density $h_{1}=c_{3}=-2c_{1}w^{2}+c_{1}(w^{1})^{2}$:%
\begin{equation}
w_{t}^{1}=2\left( \frac{(w^{1})^{2}}{2}-w^{2}\right) _{x},\text{ \ \ }%
w_{t}^{2}=2c_{1,x},\text{ \ \ }c_{1,t}=2\left( c_{1}w^{1}\right) _{x}.
\label{gora}
\end{equation}

Under the substitution $r=\Lambda p+1$, (\ref{quasi}) leads to%
\begin{equation}
\Lambda ^{2}(p^{3}-p^{2}w^{1}-2pw^{2}-w^{3})+\epsilon \Lambda
(3pp_{x}-pw_{x}^{1}-w^{1}p_{x}-w_{x}^{2})+w^{1}-\epsilon
w_{x}^{1}-p+\epsilon ^{2}p_{xx}=0.  \label{curv}
\end{equation}%
Then (\ref{ellip}) becomes (here $\Lambda ^{-2}\rightarrow 2\tilde{\lambda}$)%
\begin{equation}
\tilde{\lambda}=\frac{p^{2}}{2}-w^{2}-\frac{\frac{1}{2}w^{3}+w^{1}w^{2}}{%
p-w^{1}}.  \label{alg}
\end{equation}%
One can verify by straightforward computation that hydrodynamic type system (%
\ref{sys}) possesses a generating function of conservation laws%
\begin{equation}
p_{t}=2\left( \frac{p^{2}}{2}-w^{2}\right) _{x}.  \label{new}
\end{equation}

\textbf{Remark}: Algebraic curve of genus zero (\ref{alg}) is nothing but a
dispersionless limit of the pseudo-differential operator%
\begin{equation}
L=\partial _{x}^{2}-2w^{2}+m\partial _{x}^{-1}n  \label{yajima}
\end{equation}%
well known in theory of the Yajima--Oikawa system (see these detail, for
instance, in \cite{MaksTsar}), generating function of conservation law
densities (\ref{new}) is common for any hydrodynamic reduction of the
dispersionless KP hierarchy including a dispersionless limit of the
Yajima--Oikawa system (see detail in \cite{maksbenney}). Thus this
hydrodynamic type system (\ref{sys}) is of the Egorov type and moreover is
associated with a particular solution%
\begin{equation*}
z(w^{1},c_{1})=\frac{1}{6}c_{1}(w^{1})^{3}+\frac{3}{4}(c_{1})^{2}-\frac{1}{2}%
(c_{1})^{2}\ln c_{1}.
\end{equation*}%
of another WDVV\ associativity equation (see detail in \cite{Dubr} and \cite%
{MaksEgor}):%
\begin{equation}
\frac{\partial ^{3}z}{\partial c_{1}\partial (w^{1})^{2}}\frac{\partial ^{3}z%
}{\partial (c_{1})^{2}\partial w^{1}}=1+\frac{\partial ^{3}z}{\partial
(c_{1})^{3}}\frac{\partial ^{3}z}{\partial (w^{1})^{3}}.  \label{other}
\end{equation}

\subsection{An Integrable Dispersive Extension}

\label{subsec:lagran}

If $\epsilon \rightarrow 0$, but $\epsilon \neq 0$, then intermediate\
dispersive system can be considered as an integrable dispersive perturbation
of hydrodynamic type system (\ref{sys}):%
\begin{equation*}
v_{t}^{1}=[(v^{1}+v^{2})(v^{1}+v^{3})+\epsilon (...)+\epsilon
^{2}(...)+...]_{x},
\end{equation*}%
\begin{equation}
v_{t}^{2}=[(v^{1}+v^{2})(v^{2}+v^{3})+\epsilon (...)+\epsilon
^{2}(...)+...]_{x},  \label{sis}
\end{equation}%
\begin{equation*}
v_{t}^{3}=[(v^{1}+v^{3})(v^{2}+v^{3})+\epsilon (...)+\epsilon
^{2}(...)+...]_{x},
\end{equation*}%
which can be written in the Hamiltonian form%
\begin{equation}
v_{t}^{i}=K^{im}\left( \frac{\delta \mathbf{H}_{2}}{\delta v^{m}}\right)
_{x},  \label{loc}
\end{equation}%
where the Hamiltonian density is (see (\ref{uno}), (\ref{due})) 
\begin{equation}
h_{2}=(1-\epsilon \partial _{x})(2-\epsilon \partial _{x})c_{1}-\epsilon
(c_{0}w^{1})_{x}=w^{3}+2w^{1}(1-\epsilon \partial _{x})^{-1}w^{2},
\label{notyet}
\end{equation}%
or in expanded form:%
\begin{equation}
h_{2}=w^{3}+2w^{1}w^{2}+2\epsilon w^{1}w_{x}^{2}+2\epsilon
^{2}w^{1}w_{xx}^{2}+2\epsilon ^{3}w^{1}w_{xxx}^{2}+...  \label{expand}
\end{equation}%
Taking into account that (\ref{trua}) can be written in the quadratic form%
\begin{equation*}
(1-\epsilon \partial _{x})c_{2}=\frac{1}{2}(c_{0})^{2}-c_{1}w^{1},
\end{equation*}%
the above intermediate\ dispersive system also can be equipped by another
local Hamiltonian structure%
\begin{equation}
w_{t}^{1}=\left( \frac{\delta \mathbf{H}_{1}}{\delta c_{1}}\right) _{x},%
\text{ \ \ }c_{0,t}=-\left( \frac{\delta \mathbf{H}_{1}}{\delta c_{0}}%
\right) _{x},\text{ \ \ }c_{1,t}=\left( \frac{\delta \mathbf{H}_{1}}{\delta
w^{1}}\right) _{x},  \label{secondo}
\end{equation}%
where the Hamiltonian density is (see (\ref{uno}), (\ref{trua}), (\ref%
{quatro}))%
\begin{equation*}
h_{1}=\frac{1}{2}(1-\epsilon \partial _{x})(2-\epsilon \partial _{x})c_{3}+%
\frac{3}{2}\epsilon (c_{0}c_{1})_{x}-\frac{1}{2}\epsilon (c_{2}w^{1})_{x}
\end{equation*}%
\begin{equation*}
=\frac{1}{2}(c_{0})^{2}w^{1}+2c_{0}c_{1}+w^{1}(1-\epsilon \partial
_{x})^{-1}\left( c_{1}w^{1}-\frac{1}{2}(c_{0})^{2}\right) +\epsilon
c_{1}c_{0,x},
\end{equation*}%
or in expanded form:%
\begin{equation*}
h_{1}=2c_{0}c_{1}+c_{1}(w^{1})^{2}+\epsilon c_{1}c_{0,x}+\epsilon \left( 
\frac{1}{2}(c_{0})^{2}-c_{1}w^{1}\right) w_{x}^{1}
\end{equation*}%
\begin{equation*}
+\epsilon ^{2}\left( c_{1}w^{1}-\frac{1}{2}(c_{0})^{2}\right)
w_{xx}^{1}+\epsilon ^{3}\left( \frac{1}{2}(c_{0})^{2}-c_{1}w^{1}\right)
w_{xxx}^{1}+...
\end{equation*}%
In flat coordinates $w^{1},c_{0},c_{1}$ of second local Hamiltonian
structure (\ref{secondo}) the intermediate\ dispersive system is nonlocal:%
\begin{equation*}
w_{t}^{1}=[2c_{0}+\epsilon c_{0,x}+w^{1}(1+\epsilon \partial
_{x})^{-1}w^{1}]_{x},
\end{equation*}%
\begin{equation*}
c_{0,t}=[-2c_{1}+\epsilon c_{1,x}+c_{0}w^{1}-c_{0}(1+\epsilon \partial
_{x})^{-1}w^{1}]_{x},
\end{equation*}%
\begin{equation*}
c_{1,t}=\left( \frac{1}{2}(c_{0})^{2}+c_{1}(1+\epsilon \partial
_{x})^{-1}w^{1}+(1-\epsilon \partial _{x})^{-1}\left( c_{1}w^{1}-\frac{1}{2}%
(c_{0})^{2}\right) \right) _{x}.
\end{equation*}%
If we introduce the field variable $g$ such that%
\begin{equation*}
w^{1}=(1+\epsilon \partial _{x})g,
\end{equation*}%
then the intermediate\ dispersive system can be written in a little bit more
simple form%
\begin{equation}
(1+\epsilon \partial _{x})g_{t}=[2c_{0}+\epsilon c_{0,x}+g(g+\epsilon
g_{x})]_{x},  \label{lok}
\end{equation}%
\begin{equation*}
c_{0,t}=[-2c_{1}+\epsilon c_{1,x}-\epsilon c_{0}g_{x}]_{x},
\end{equation*}%
\begin{equation*}
(1-\epsilon \partial _{x})c_{1,t}=[(2c_{1}-\epsilon c_{1,x})g-\epsilon
c_{0}c_{0,x}]_{x}.
\end{equation*}%
Its Hamiltonian structure is%
\begin{equation*}
(1+\epsilon \partial _{x})g_{t}=\left( \frac{\delta \mathbf{H}_{1}}{\delta
c_{1}}\right) _{x},\text{ \ \ }c_{0,t}=-\left( \frac{\delta \mathbf{H}_{1}}{%
\delta c_{0}}\right) _{x},\text{ \ \ }(1-\epsilon \partial
_{x})c_{1,t}=\left( \frac{\delta \mathbf{H}_{1}}{\delta g}\right) _{x},
\end{equation*}%
where%
\begin{equation*}
h_{1}=c_{1}[2c_{0}+\epsilon c_{0,x}+g^{2}+\epsilon gg_{x}]+\frac{1}{2}%
\epsilon g_{x}(c_{0})^{2}.
\end{equation*}%
A corresponding local Lagrangian representation is%
\begin{equation*}
S_{1}=\int [\frac{1}{2}(G_{x}-\epsilon ^{2}G_{xxx})Q_{t}+\frac{1}{2}%
(Q_{x}-\epsilon ^{2}Q_{xxx})G_{t}-\frac{1}{2}F_{x}F_{t}-h_{1}]dxdt,
\end{equation*}%
where we introduced functions $Q,F,G$ such that%
\begin{equation*}
c_{1}=(1+\epsilon \partial _{x})Q_{x},\text{ \ \ }g=G_{x},\text{ \ }%
c_{0}=F_{x}.
\end{equation*}

Since any conservation law density is determined up to a total derivative,
the Hamiltonian density (\ref{expand}) can be also written in the expanded
form%
\begin{equation*}
h_{2}=w^{3}+2w^{1}w^{2}-2\epsilon w^{2}w_{x}^{1}+2\epsilon
^{2}w^{2}w_{xx}^{1}-2\epsilon ^{3}w^{2}w_{xxx}^{1}+...
\end{equation*}%
This means that this Hamiltonian density in a compact form becomes (cf. (\ref%
{notyet}))%
\begin{equation*}
h_{2}=w^{3}+2w^{2}(1+\epsilon \partial _{x})^{-1}w^{1}.
\end{equation*}%
Taking into account again (\ref{vieta}), the Hamiltonian density takes the
form%
\begin{equation*}
h_{2}=v^{1}v^{2}v^{3}-(v^{1}v^{2}+v^{1}v^{3}+v^{2}v^{3})(1+\epsilon \partial
_{x})^{-1}(v^{1}+v^{2}+v^{3}).
\end{equation*}%
Then introducing differential substitutions%
\begin{equation*}
v^{k}=s^{k}+\epsilon s_{x}^{k},
\end{equation*}%
the Hamiltonian density reduces to the local expression%
\begin{equation}
h_{2}=(s^{1}+\epsilon s_{x}^{1})(s^{2}+\epsilon s_{x}^{2})(s^{3}+\epsilon
s_{x}^{3})  \label{yok}
\end{equation}%
\begin{equation*}
-(s^{1}+s^{2}+s^{3})[(s^{1}+\epsilon s_{x}^{1})(s^{2}+\epsilon
s_{x}^{2})+(s^{1}+\epsilon s_{x}^{1})(s^{3}+\epsilon
s_{x}^{3})+(s^{2}+\epsilon s_{x}^{2})(s^{3}+\epsilon s_{x}^{3})].
\end{equation*}%
Then local Hamiltonian structure (\ref{loc}) transforms into nonlocal:%
\begin{equation*}
s_{t}^{i}=K^{im}(1-\epsilon ^{2}\partial _{x}^{2})^{-1}\left( \frac{\delta 
\mathbf{H}_{2}}{\delta s^{m}}\right) _{x}.
\end{equation*}%
A corresponding intermediate\ dispersive system takes the following
non-evolutionary form%
\begin{equation}
s_{t}^{i}-\epsilon ^{2}s_{xxt}^{i}=K^{im}\left( \frac{\delta \mathbf{H}_{2}}{%
\delta s^{m}}\right) _{x},  \label{nonevo}
\end{equation}%
where (see (\ref{key}) and (\ref{yok})), for instance,%
\begin{equation*}
K^{1m}\frac{\delta \mathbf{H}_{2}}{\delta s^{m}}=(s^{1}-s^{2})(s^{1}-s^{3})
\end{equation*}%
\begin{equation*}
+\frac{\epsilon }{2}%
[(-2s^{1}+s^{2}+s^{3})s_{x}^{1}+(s^{3}-s^{1})s_{x}^{2}+(s^{2}-s^{1})s_{x}^{3}]
\end{equation*}%
\begin{equation*}
+\frac{\epsilon ^{2}}{2}%
[-(2s^{1}+s^{2}+s^{3})s_{xx}^{1}+(s^{1}-s^{3})s_{xx}^{2}+(s^{1}-s^{2})s_{xx}^{3}-2(s_{x}^{1})^{2}]
\end{equation*}%
\begin{equation*}
+\frac{\epsilon ^{3}}{2}%
[(s_{x}^{2}+s_{x}^{3})s_{xx}^{1}+(s_{x}^{1}-s_{x}^{3})s_{xx}^{2}+(s_{x}^{1}-s_{x}^{2})s_{xx}^{3}],
\end{equation*}%
while two other expressions $K^{2m}\frac{\delta \mathbf{H}_{2}}{\delta s^{m}}
$ and $K^{3m}\frac{\delta \mathbf{H}_{2}}{\delta s^{m}}$ can be obtained by
a cyclic permutation of indices. Thus the intermediate\ dispersive system
also has a local Lagrangian representation (see (\ref{nonevo}))%
\begin{equation*}
S_{2}=\int \left( \frac{1}{2}K_{im}(\varphi _{x}^{i}-\epsilon ^{2}\varphi
_{xxx}^{i})\varphi _{t}^{m}-h_{2}\right) dxdt,
\end{equation*}%
where $\varphi ^{k}$ are potentials such that $s^{k}=\varphi _{x}^{k}$ and $%
K_{im}$ is determined by (\ref{key}).

In field variables $w^{k}$ the first local Hamiltonian operator is precisely
(\ref{i}):%
\begin{equation}
\hat{A}_{1}=%
\begin{pmatrix}
-\frac{3}{2}\partial _{x}^{{}} & \frac{1}{2}\partial _{x}^{{}}w^{1} & 
\partial _{x}^{{}}w^{2} \\ 
\frac{1}{2}w^{1}\partial _{x}^{{}} & \frac{1}{2}(\partial
_{x}^{{}}w^{2}+w^{2}\partial _{x}^{{}}) & \frac{3}{2}w^{3}\partial
_{x}^{{}}+w_{x}^{3} \\ 
w^{2}\partial _{x}^{{}} & \frac{3}{2}\partial _{x}^{{}}w^{3}-w_{x}^{3} & 
[(w^{2})^{2}-w^{1}w^{3}]\partial _{x}^{{}}+\partial
_{x}^{{}}[(w^{2})^{2}-w^{1}w^{3}]%
\end{pmatrix}%
.  \label{dubl}
\end{equation}%
One can recalculate a second local Hamiltonian operator in the same
coordinates $w^{k}$, taking into account (\ref{uno}) and differential
substitutions (\ref{uno}), (\ref{due}), (\ref{trua}). In this case a first
order local Hamiltonian operator written in field variables $%
w^{1},c_{0},c_{1}$ becomes nonhomogeneous local third order Hamiltonian
operator written in field variables $w^{k}$:%
\begin{equation}
w_{t}^{1}=(2\partial _{x}+3\epsilon \partial _{x}^{2}+\epsilon ^{2}\partial
_{x}^{3})\frac{\delta \mathbf{H}_{1}}{\delta w^{3}},  \label{extend}
\end{equation}%
\begin{equation*}
w_{t}^{2}=-\partial _{x}\frac{\delta \mathbf{H}_{1}}{\delta w^{2}}%
+2(w_{x}^{1}+w^{1}\partial _{x})\frac{\delta \mathbf{H}_{1}}{\delta w^{3}}%
+\epsilon \left( -2w_{xx}^{1}-3w_{x}^{1}\partial _{x}-w^{1}\partial
_{x}^{2}\right) \frac{\delta \mathbf{H}_{1}}{\delta w^{3}}
\end{equation*}%
\begin{equation*}
+\epsilon ^{2}\left( \partial _{x}^{3}\frac{\delta \mathbf{H}_{1}}{\delta
w^{2}}-(w_{xx}^{1}\partial _{x}+2w_{x}^{1}\partial _{x}^{2}+w^{1}\partial
_{x}^{3})\frac{\delta \mathbf{H}_{1}}{\delta w^{3}}\right) ,
\end{equation*}%
\begin{equation*}
w_{t}^{3}=2\partial _{x}\frac{\delta \mathbf{H}_{1}}{\delta w^{1}}%
+2w^{1}\partial _{x}\frac{\delta \mathbf{H}_{1}}{\delta w^{2}}%
-(4w^{1}w_{x}^{1}+4w_{x}^{2}+(8w^{2}+4(w^{1})^{2})\partial _{x})\frac{\delta 
\mathbf{H}_{1}}{\delta w^{3}}
\end{equation*}%
\begin{equation*}
+\epsilon \left( -3\partial _{x}^{2}\frac{\delta \mathbf{H}_{1}}{\delta w^{1}%
}+(w^{1}\partial _{x}^{2}-w_{x}^{1}\partial _{x})\frac{\delta \mathbf{H}_{1}%
}{\delta w^{2}}%
+(2w^{1}w_{xx}^{1}+2w_{xx}^{2}+2(w_{x}^{1})^{2}+(4w^{1}w_{x}^{1}+4w_{x}^{2})%
\partial _{x})\frac{\delta \mathbf{H}_{1}}{\delta w^{3}}\right) 
\end{equation*}%
\begin{equation*}
+\epsilon ^{2}\left( \partial _{x}^{3}\frac{\delta \mathbf{H}_{1}}{\delta
w^{1}}-(w_{x}^{1}\partial _{x}^{2}+w^{1}\partial _{x}^{3})\frac{\delta 
\mathbf{H}_{1}}{\delta w^{2}}\right) 
\end{equation*}%
\begin{equation*}
+\epsilon ^{2}[(w_{xx}^{2}+w^{1}w_{xx}^{1}+(w_{x}^{1})^{2})\partial
_{x}+(3w^{1}w_{x}^{1}+3w_{x}^{2})\partial
_{x}^{2}+((w^{1})^{2}+2w^{2})\partial _{x}^{3}]\frac{\delta \mathbf{H}_{1}}{%
\delta w^{3}}.
\end{equation*}%
Intermediate\ dispersive system (\ref{sis}) in field variables $w^{k}$ takes
the form (see (\ref{dubl}))%
\begin{equation*}
w_{t}^{1}=[(w^{1})^{2}-2w^{2}-\epsilon (w^{1}w_{x}^{1}+3w_{x}^{2})+\epsilon
^{2}(w^{1}w_{xx}^{1}-3w_{xx}^{2})-\epsilon
^{3}(w^{1}w_{xxx}^{1}+3w_{xxx}^{2})+...]_{x},
\end{equation*}%
\begin{equation*}
w_{t}^{2}=[2w^{1}w^{2}+w^{3}+\epsilon
(w^{1}w_{x}^{2}-2w^{2}w_{x}^{1})+\epsilon
^{2}(w^{1}w_{xx}^{2}-w_{x}^{1}w_{x}^{2}+2w^{2}w_{xx}^{1})
\end{equation*}%
\begin{equation*}
+\epsilon
^{3}(w^{1}w_{xxx}^{2}+w_{x}^{2}w_{xx}^{1}-w_{x}^{1}w_{xx}^{2}-2w^{2}w_{xxx}^{1})+...]_{x},
\end{equation*}%
\begin{equation*}
w_{t}^{3}=4w^{2}w_{x}^{2}+2w^{3}w_{x}^{1}+\epsilon
(2w^{2}w_{xx}^{2}-w_{x}^{1}w_{x}^{3}-3w^{3}w_{xx}^{1})
\end{equation*}%
\begin{equation*}
+\epsilon
^{2}(2w^{2}w_{xxx}^{2}+w_{x}^{3}w_{xx}^{1}+3w^{3}w_{xxx}^{1})+\epsilon
^{3}(2w^{2}w_{xxxx}^{2}-w_{x}^{3}w_{xxx}^{1}-3w^{3}w_{xxxx}^{1})+...,
\end{equation*}%
where the Hamiltonian density%
\begin{equation*}
h_{1}=({w^{1})}^{3}{w^{2}}-2{w^{1}(}{w^{2})}^{2}+\frac{1}{2}({w^{1})}^{2}{%
w^{3}}-{w^{2}}{w^{3}}+\epsilon \left( 2{(}{w^{2})}^{2}{w_{x}^{1}}+2({w^{1})}%
^{3}{w_{x}^{2}}+{(w^{1})}^{2}{w_{x}^{3}}\right) +...
\end{equation*}

\section{The Yajima--Oikawa System and its Dispersionless Limit}

\label{sec:oikawa}

The remarkable Yajima--Oikawa system is associated with linear equation (\ref%
{yajima}):%
\begin{equation*}
\psi _{xx}-2w^{2}\psi +m\partial _{x}^{-1}n\psi =\lambda \psi ,
\end{equation*}%
which can be written as a third order linear ordinary differential equation%
\begin{equation*}
\psi _{xxx}-\frac{m_{x}}{m}\psi _{xx}-2w^{2}\psi _{x}+\left(
-2w_{x}^{2}+2w^{2}\frac{m_{x}}{m}\right) \psi +nm\psi =\lambda \left( \psi
_{x}-\frac{m_{x}}{m}\psi \right)
\end{equation*}%
or in the factorized form%
\begin{equation*}
(\partial _{x}-v^{1})(\partial _{x}-v^{2})(\partial _{x}-v^{3})\psi =\lambda
(\partial _{x}-w^{1})\psi .
\end{equation*}%
Substitution (\ref{r}) leads to the nonlinear ordinary differential equation
of second order%
\begin{equation*}
r_{xx}+3rr_{x}+r^{3}-w^{1}(r_{x}+r^{2})-2w^{2}r-w^{3}=\lambda (r-w^{1}),
\end{equation*}%
where we introduced new field variables $w^{1}=(\ln
m)_{x},w^{3}=2w_{x}^{2}-2w^{1}w^{2}-nm$.

Now we incorporate a parameter $\epsilon $ changing $\partial
_{x}\rightarrow \epsilon \partial _{x}$, and consider the equation%
\begin{equation*}
\epsilon ^{2}r_{xx}+3\epsilon rr_{x}+r^{3}-w^{1}(\epsilon
r_{x}+r^{2})-2w^{2}r-w^{3}=\lambda (r-w^{1}),
\end{equation*}%
whose unknown functions $w^{k}$ can be expressed via roots $v^{k}$ (cf. (\ref%
{vieta})):%
\begin{equation*}
w^{1}=v^{1}+v^{2}+v^{3},\text{ \ \ }%
-2w^{2}=v^{1}v^{2}+v^{1}v^{3}+v^{2}v^{3}-\epsilon (v_{x}^{2}+2v_{x}^{3}),
\end{equation*}%
\begin{equation*}
w^{3}=v^{1}v^{2}v^{3}-\epsilon
(v^{1}v_{x}^{3}+v^{2}v_{x}^{3}+v^{3}v_{x}^{2})+\epsilon ^{2}v_{xx}^{3}.
\end{equation*}

Below we present main results of this Section and omit all computations,
because they are similar to the previous Sections.

\textbf{Proposition}: \textit{The evolutionary system }(cf. (\ref{gora}))%
\begin{equation}
w_{t}^{1}=2\left( \frac{(w^{1})^{2}}{2}-w^{2}+\frac{\epsilon }{2}%
w_{x}^{1}\right) _{x},\text{ \ }w_{t}^{2}=2c_{1,x},\text{ \ }c_{1,t}=2\left(
c_{1}w^{1}-\frac{\epsilon }{2}c_{1,x}\right) _{x}  \label{w}
\end{equation}%
\textit{is a bi-Hamiltonian system}.

Indeed $w^{1},w^{2},c_{1}$ are flat coordinates of the second local
Hamiltonian structure, i.e. (cf. (\ref{egor}))%
\begin{equation*}
w_{t}^{1}=\left( \frac{\delta \mathbf{H}_{1}}{\delta c_{1}}\right) _{x},%
\text{ \ }w_{t}^{2}=-\left( \frac{\delta \mathbf{H}_{1}}{\delta w^{2}}%
\right) _{x},\text{ \ }c_{1,t}=\left( \frac{\delta \mathbf{H}_{1}}{\delta
w^{1}}\right) _{x},
\end{equation*}%
where the Hamiltonian density is $h_{1}=c_{1}[(w^{1})^{2}-2w^{2}+\epsilon
w_{x}^{1}]$, and $c_{1}=\frac{1}{2}w^{3}+w^{1}w^{2}-\epsilon w_{x}^{2}$ is a
conservation law density.

In flat coordinates $v^{k}$ of the first local Hamiltonian structure%
\begin{equation*}
v_{t}^{i}=K^{im}\left( \frac{\delta \mathbf{H}_{2}}{\delta v^{m}}\right) _{x}
\end{equation*}%
evolutionary system (\ref{w}) takes the form%
\begin{eqnarray*}
v_{t}^{1} &=&[{(v^{1}}+{v^{2})(v^{1}}+{v^{3})}+\epsilon (v^{1}+{v^{2})}_{x}{]%
}_{x}, \\
v_{t}^{2} &=&[{(v^{1}}+{v^{2})(v^{2}}+{v^{3})]}_{x}, \\
v_{t}^{3} &=&[{(v^{1}}+{v^{3})(v^{2}}+{v^{3})}-\epsilon ({v^{2}}+{v^{3})}_{x}%
{]}_{x},
\end{eqnarray*}%
where the Hamiltonian density is%
\begin{equation*}
h_{2}=2w^{1}w^{2}+w^{3}
\end{equation*}%
\begin{equation*}
=v^{1}v^{2}v^{3}-(v^{1}+v^{2}+v^{3})(v^{1}v^{2}+v^{1}v^{3}+v^{2}v^{3})+%
\epsilon (v^{1}v_{x}^{2}+(v^{1}+v^{2})v_{x}^{3}).
\end{equation*}

\section{Another Dispersionful Version of the WDVV associativity System}

\label{sec:another}

In this paper we dealt with the first \textquotedblleft
half\textquotedblright\ of linear spectral problem (\ref{a}). However
completely the same set of computations can be made for the second
\textquotedblleft half\textquotedblright\ of this linear spectral problem (%
\ref{a}). Indeed the matrix system%
\begin{equation*}
\mathbf{\psi }_{y}=\lambda 
\begin{pmatrix}
0 & 0 & 1 \\ 
a^{3} & a^{2} & 0 \\ 
(a^{2})^{2}-a^{1}a^{3} & a^{3} & 0%
\end{pmatrix}%
\mathbf{\psi }
\end{equation*}%
can be written as a single ordinary differential equation of third order%
\begin{equation*}
\psi _{yyy}-\left( \lambda a^{2}+\frac{a_{y}^{3}}{a^{3}}\right) \psi
_{yy}+\lambda ^{2}[a^{1}a^{3}-(a^{2})^{2}]\psi _{y}+\lambda ^{2}\left(
a^{3}a_{y}^{1}-2a^{2}a_{y}^{2}+\frac{(a^{2})^{2}}{a^{3}}a_{y}^{3}+\lambda
\lbrack (a^{2})^{3}-a^{1}a^{2}a^{3}-(a^{3})^{2}]\right) \psi =0,
\end{equation*}%
where%
\begin{equation*}
\psi _{2}=\frac{1}{\lambda }\psi _{y},\text{ \ }\psi _{1}=\frac{1}{\lambda
^{2}a^{3}}\psi _{yy}+\left( a^{1}-\frac{(a^{2})^{2}}{a^{3}}\right) \psi .
\end{equation*}%
Under the scaling transformation (cf. (\ref{z}))%
\begin{equation}
y=\epsilon (e^{t/\epsilon }-1),  \label{y}
\end{equation}%
this third order equation becomes%
\begin{equation*}
\psi _{ttt}-\left( \lambda w^{2}+\frac{3}{2\epsilon }+\frac{w_{t}^{3}}{w^{3}}%
\right) \psi _{tt}+\left( \lambda ^{2}[w^{1}w^{3}-(w^{2})^{2}]+\frac{1}{%
\epsilon }\lambda w^{2}+\frac{1}{2\epsilon ^{2}}+\frac{1}{\epsilon }\frac{%
w_{t}^{3}}{w^{3}}\right) \psi _{t}
\end{equation*}%
\begin{equation*}
+\lambda ^{2}\left( w^{3}w_{t}^{1}-2w^{2}w_{t}^{2}+\frac{(w^{2})^{2}}{w^{3}}%
w_{t}^{3}-\frac{1}{2\epsilon }w^{1}w^{3}+\frac{1}{2\epsilon }%
(w^{2})^{2}+\lambda \lbrack (w^{2})^{3}-w^{1}w^{2}w^{3}-(w^{3})^{2}]\right)
\psi =0,
\end{equation*}%
where we introduced new field variables $w^{k}$ such that%
\begin{equation*}
a^{1}=w^{1}e^{-t/(2\epsilon )},\text{ \ }a^{2}=w^{2}e^{-t/\epsilon },\text{
\ \ }a^{3}=w^{3}e^{-3t/(2\epsilon )}.
\end{equation*}

In a high-frequency limit corresponding WDVV associativity system also has a
bi-Hamiltonian structure (see detail in \cite{Nutku}). One can completely
repeat the computations made above in this paper and derive another
dispersionful version of an intermediate system whose high-frequency limit
must coincide with original WDVV associativity system up to change of
independent variables ($y\leftrightarrow z$).%
\begin{equation*}
\epsilon ^{2}r_{tt}+3\epsilon rr_{t}+r^{3}-\left( \Lambda w^{2}+\frac{3}{2}%
+\epsilon \frac{w_{t}^{3}}{w^{3}}\right) (\epsilon r_{t}+r^{2})+\left(
\Lambda ^{2}[w^{1}w^{3}-(w^{2})^{2}]+\Lambda w^{2}+\frac{1}{2}+\epsilon 
\frac{w_{t}^{3}}{w^{3}}\right) r
\end{equation*}%
\begin{equation*}
+\Lambda ^{2}\left[ \frac{1}{2}(w^{2})^{2}-\frac{1}{2}w^{1}w^{3}+\epsilon
\left( w^{3}w_{t}^{1}-2w^{2}w_{t}^{2}+\frac{(w^{2})^{2}}{w^{3}}%
w_{t}^{3}\right) \right] +\Lambda
^{3}[(w^{2})^{3}-w^{1}w^{2}w^{3}-(w^{3})^{2}]=0,
\end{equation*}%
The substitution $r=\Lambda p+\frac{1}{2}$ yields (cf. (\ref{curv}))%
\begin{equation*}
\Lambda
^{3}[p^{3}-w^{2}p^{2}+[w^{1}w^{3}-(w^{2})^{2}]p+(w^{2})^{3}-w^{1}w^{2}w^{3}-(w^{3})^{2}]
\end{equation*}%
\begin{equation*}
+\epsilon \Lambda ^{2}\left[ w^{3}w_{t}^{1}-2w^{2}w_{t}^{2}+\frac{(w^{2})^{2}%
}{w^{3}}w_{t}^{3}+3pp_{t}-w^{2}p_{t}-\frac{w_{t}^{3}}{w^{3}}p^{2}\right]
\end{equation*}%
\begin{equation*}
+\Lambda \left( \frac{1}{4}w^{2}-\frac{1}{4}p+\epsilon ^{2}p_{tt}-\epsilon
^{2}\frac{w_{t}^{3}}{w^{3}}p_{t}\right) +\frac{1}{4}\epsilon \frac{w_{t}^{3}%
}{w^{3}}=0.
\end{equation*}%
Then in a dispersionless limit one can obtain again (cf. (\ref{alg}))%
\begin{equation*}
\bar{\lambda}=p^{2}+w^{1}w^{3}-(w^{2})^{2}-\frac{(w^{3})^{2}}{p-w^{2}},
\end{equation*}%
where $\bar{\lambda}=(2\Lambda )^{-2}$. This means that this dispersionless
limit also is associated with the Yajima--Oikawa hierarchy.

\section{Conclusion}

\label{sec:final}

In this paper we investigated the intermediate dispersive system, which
possesses two limits:

1. a high-frequency limit, nondiagonalizable hydrodynamic type system (\ref%
{ha})%
\begin{equation*}
u_{y}^{1}=\frac{1}{2}(u^{2}u^{3}-u^{1}u^{2}-u^{1}u^{3})_{z},\text{ \ }%
u_{y}^{2}=\frac{1}{2}(u^{1}u^{3}-u^{1}u^{2}-u^{2}u^{3})_{z},\text{ \ }%
u_{y}^{3}=\frac{1}{2}(u^{1}u^{2}-u^{1}u^{3}-u^{2}u^{3})_{z}
\end{equation*}%
integrable by the inverse scattering transform;

2. a dispersionless limit, semi-Hamiltonian hydrodynamic type system (\ref%
{sys})%
\begin{equation*}
v_{t}^{1}=[(v^{1}+v^{2})(v^{1}+v^{3})]_{x},\text{ \ }%
v_{t}^{2}=[(v^{1}+v^{2})(v^{2}+v^{3})]_{x},\text{ \ }%
v_{t}^{3}=[(v^{1}+v^{3})(v^{2}+v^{3})]_{x}
\end{equation*}%
integrable by the generalized hodograph method (see detail in \cite{Tsar}).

Introducing the potential function $F$ such that (see (\ref{gora}))%
\begin{equation*}
w^{2}=F_{xx},\text{ \ }c_{1}=\frac{1}{2}F_{xt},\text{ \ }w^{1}=\frac{F_{tt}}{%
2F_{xt}}
\end{equation*}%
this semi-Hamiltonian hydrodynamic type system can be written as a single
equation of third order (cf. (\ref{sing}))%
\begin{equation*}
F_{ttt}=2\frac{F_{tt}}{F_{xt}}F_{xtt}-\frac{F_{tt}^{2}}{F_{xt}^{2}}%
F_{xxt}-4F_{xt}F_{xxx}.
\end{equation*}

The WDVV associativity system (\ref{ha}) admits two different dispersionful
extensions based on simple transformations (\ref{z}) and (\ref{y}). Their
dispersionless limits coincide with a dispersionless limit of the
Yajima--Oikawa system (\ref{sys}).

The intermediate dispersive system (see (\ref{non}), here we denote $\xi
=\epsilon ^{-1}$)%
\begin{equation}
(\partial _{x}+\xi )\eta _{t}=(q_{0,x}+\xi (2q_{0}+\eta \eta _{x})+\xi
^{2}\eta ^{2})_{x},\text{ \ \ }q_{0,t}=(q_{1,x}-q_{0}\eta _{x}-2\xi
q_{1})_{x},  \label{ksi}
\end{equation}%
\begin{equation*}
(\partial _{x}-\xi )q_{1,t}=(q_{0}q_{0,x}+\xi \eta q_{1,x}-2\xi ^{2}\eta
q_{1})_{x}
\end{equation*}%
reduces to original WDVV associativity equation (\ref{sing}) in a
high-frequency limit ($\xi \rightarrow 0$), where $\eta =f_{xx},q_{0}=f_{xt}$
and the function $q_{1}$ is determined by its first derivatives (the
compatibility condition $(q_{1,x})_{t}=(q_{1,t})_{x}$ yields again (\ref%
{sing})):%
\begin{equation*}
q_{1,x}=f_{tt}+f_{xt}f_{xxx},\text{ \ }q_{1,t}=f_{xt}f_{xxt}.
\end{equation*}%
The intermediate dispersive system is bi-Hamiltonian and both corresponding
Lagrangian representations (\ref{k}) and (\ref{s}) are local.

In Dubrovin's approach (the so-called integrable higher dispersive
corrections) theory of WDVV associativity equations plays a very important
role. Any solution of (\ref{sing}) leads to two three-component commuting
hydrodynamic type systems (see detail in \cite{Dubr} and \cite{MaksEgor}),
which are Hamiltonian and Egorov (see detail in \cite{MaksTsar}). Integrable
dispersive corrections of these hydrodynamic type systems simultaneously
allow to construct integrable dispersive correction for WDVV associativity
system (\ref{wdvv}). In our paper we used an alternative strategy: we
constructed an integrable perturbation (\ref{ksi}). If $\xi \rightarrow 0$
(i.e. a high-frequency limit), we came back to (\ref{wdvv}); if $\xi
\rightarrow \infty $ (i.e. a dispersionless limit), we obtained a single
three-component hydrodynamic type system (\ref{sys}), which is
bi-Hamiltonian and Egorov. This system is determined by a particular
solution of another WDVV associativity equation (see (\ref{other}) and cf. (%
\ref{sing}))%
\begin{equation*}
f_{xxt}f_{xtt}=1+f_{xxx}f_{ttt}.
\end{equation*}%
Both WDVV associativity equations are connected with each other by a special
reciprocal transformation (see \cite{FM}). If $\xi $ is arbitrary (assume
for simplicity that $\xi $ is small), then intermediate dispersive system (%
\ref{ksi}) can be interpreted as an integrable system describing integrable
corrections (in any order with respect to $\xi $) of Hamiltonian and Egorov
three-component hydrodynamic type systems. However these integrable
corrections should have another interpretation in comparison with Dubrovin's
approach. This separate investigation should be made somewhere else.

If parameter $\epsilon $ is small, intermediate dispersive system (\ref{lok}%
) has a pair of Hamiltonian structures (\ref{dubl}), (\ref{extend}), which
become first order Hamiltonian structures of Dubrovin--Novikov type (see 
\cite{DN}) in a dispersionless limit. If parameter $\xi $ is small,
intermediate dispersive system (\ref{ksi}) has the same pair of Hamiltonian
structures, but the second Hamiltonian operator reduces to third order
Hamiltonian operator of Dubrovin--Novikov type (see \cite{DN2}) in a
high-frequency limit (see again \cite{FGMN}).

\section*{Acknowledgements}

Authors thank Oleksandr Chvartatskyi, Eugene Ferapontov and Folkert
Muller-Hoissen for numerous discussions and for important remarks.

MVP would like to thank the Mathematical Institute at University of G\"{o}%
ttingen for the hospitality during his visits. MVP's work was partially
supported by the grant of Presidium of RAS \textquotedblleft Fundamental
Problems of Nonlinear Dynamics\textquotedblright\ and by the RFBR grant
14-01-00012. N.M.S. is supported by the Marie Curie Actions Intra-European
fellowship HYDRON (FP7-PEOPLE-2012-IEF, Project number 332136).

\end{document}